          [2001/04/25 1.1 (PWD)]
\documentclass[a4paper,twocolumn]{esapub} 
\usepackage{natbib}

\title{2-100 keV Spectrum of an Actively Star Forming Galaxy}

\author{Massimo Persic}
\affil{INAF/Osservatorio Astronomico di Trieste, via G.B.Tiepolo 11, 
I-34131 Trieste, Italy ({\tt persic@ts.astro.it})}
\author{Yoel Rephaeli}
\affil{School of Physics and Astronomy, Tel Aviv University, 
Tel Aviv 69978, Israel ({\tt yoelr@wise1.tau.ac.il}); and 
CASS, University of California, San Diego, La Jolla, CA 92093, USA ({\tt yoelr@mamacass.ucsd.edu})}

\begin{document}


\maketitle

\begin{abstract}
We compute the synthetic 2-100 keV spectrum of an actively
star forming galaxy. To this aim we use a luminosity function of point
sources appropriate for starbursts (based on {\it Chandra} data), as well
as the types of stellar sources and their corresponding spectra. Our
estimates indicate that a Compton spectral component - resulting from
scattering of SN-accelerated electrons by ambient FIR and CMB photons -
could possibly be detected, in deep INTEGRAL observations of nearby
starburst galaxies, only if there is a break in the spectra of the
brightest ($L \geq 2 \times 10^{38}$ erg s$^{-1}$) point sources. 
\end{abstract}

\section{Introduction}

Star formation (SF) leads to X-ray emission on various spatial and
temporal scales, including emission from O stars, X-ray binaries, 
supernovae and their remnants (SNRs), and galactic-scale 
emission from diffuse hot gas, and Compton scattering of FIR \& 
CMB photons by SN-accelerated electrons. Integrated spectra of 
actively SF galaxies may show all these components as well as 
emission from an active nucleus (Rephaeli et al. 1991, 1995).

Persic \& Rephaeli (2002; hereafter, PR02) have quantitatively 
assessed the roles of the various X-ray emission mechanisms in 
SFGs. They have used an equilibrium stellar-population synthesis 
model of the Galactic population of X-ray binaries to deduce 
birthrates for interacting binaries; these, combined with 
estimates of the duration of the X-ray bright phase, have allowed 
PR02 to make realistic estimates of the relative (Galactic) 
abundances of high-mass and low-mass X-ray binaries (HMXBs and 
LMXBs, in which the optical companions are main-sequence stars 
with $M>8\,M_\odot$ and $M<M_\odot$, respectively). The abundance 
of SNRs (both Type II and Ia) was also consistently estimated. 
From the literature PR02 derived typical spectra for these source 
classes. The spectral properties and relative abundances of the 
various classes of stellar sources allowed PR02 to calculate the 
composite X-ray spectrum arising from a stellar population of 
Galactic composition.

In addition to such "stellar" components, PR02 considered 
two diffuse, non-stellar contributions to a starburst (SB) 
X-ray spectrum. 

\noindent
{\it (i)} Thermal emission from the regions of interaction between 
SN-powered outgoing galactic wind and denser ambient ISM, mostly at 
energies $\leq$1 keV, and thus of little concern to us here.

\noindent
{\it (ii)} Non-thermal emission due to Compton scattering of the 
SN-accelerated relativistic radio-emitting electrons off the FIR 
and CMB radiation fields. The high SN rate in a SB is bound to 
yield high relativistic electron densities since SN shocks are 
known to be primary sites of cosmic ray acceleration. Moreover, 
in a SB environment the mean FIR energy density can be $\sim 10$ 
times higher than that of the CMB. Compton scattering of a $100 
\mu$ photon by a $\sim$1 GeV electron boosts the energy of the 
photon to $\sim$10 keV. A high density of electrons in a SB would 
obviously be reflected also in enhanced radio emission. If the 
same electron population produces both the radio and hard X-ray 
emission, then both spectra are expected to have {\it roughly} 
the same power-law index (Rephaeli 1979; Goldshmidt \& Rephaeli 
(1995). Compton X-ray emission gauges the current SFR since the 
electron Compton-synchrotron lifetime is approximately $\leq$10$^8$ 
yr (i.e. comparable to or shorter than a typical starburst duration).

PR02 determined that the $\sim$2-15 keV emission is dominated by 
LMXBs; this is not surprising, given that the synthetic spectrum 
was calibrated by that of a normal, quietly star forming galaxy like 
ours. Due to the long times between the formation and the onset of 
the X-ray phase (the optical companion is a subsolar main-sequence 
star) in LMXBs, their emission traces the higher, {\it past} 
Galactic SFR, and not the lower, {\it current} SFR. Thus in order 
to obtain an estimate of the current SFR, the spectral contribution 
of LMXBs should not be included. As noted by PR02, at energies of 
approximately $\geq$30 keV (relevant for INTEGRAL) the SB spectrum 
is expected to be dominated by the Compton component. 

Recent observations of nearby galaxies -- mostly with the {\it Chandra} 
satellite -- have revealed several X-ray point sources (XPs) with 
2-10 keV luminosities in the range $2 \times 10^{38}$ erg s$^{-1} 
\leq L \leq 10^{39}$ erg s$^{-1}$ (hereafter dubbed Very 
Luminous Sources: VLXs), and a few Ultra-Luminous Sources (ULXs), 
with $L \geq 10^{39}$ erg s$^{-1}$ (Fabbiano \& White 2003)
        \footnote{These ranges correspond to Eddington luminosities 
        for spherical accretion on, respectively, 2 and $8\,M_\odot$ 
	BHs. This is the range of BH masses that are thought to be 
	attainable via ordinary stellar evolution. ULXs are called 
	'super-Eddington sources'.}.
Although their nature is not yet known, VLXs and ULXs are observed to be
directly linked to current star formation activity (e.g., Zezas
et al. 2002), and when identified, the optical counterpart of ULXs are O
stars (Liu et al. 2002), thus establishing a link between HMXBs and ULXs.
The spectra of VLXs and ULXs are characterized both as power-law in the
range of interest, with index $\Gamma \sim 2$ and $\sim$1.3, respectively. 
        \footnote{Broadly speaking, the VLXs and ULXs could, in fact,
        describe different luminosity/spectral states of BH X-ray binaries
        (BHXBs) as a function of the mass infall rate ($\dot M$) onto the BH,
        while the other relevant parameters are constrained within a much
        smaller range (e.g., BH mass $M \sim 5-10 M_\odot$, efficiency
        $\eta \sim 0.3 - 0.6$). In fact, qualitatively speaking, the inverse
        correlation between luminosity and spectral index (i.e., flatter
        spectra for higher luminosities) can be understood within the
        accretion model devised to explain the high X-ray luminosities of
        accretion-powered binary systems. The increase in luminosity is
        driven by an increase of $\dot M$, which in turn means a piling
	up of material around the emission region that leads to a higher
        Compton optical depth, and hence to a lower value of $\Gamma$ i.e.
        a flatter spectrum. Namely, if a source is embedded in a plasma
        cloud of optical depth $\tau$ and temperature $T_{\rm e}$, the
        spectral index $\Gamma$ of the high-frequency wing of the spectrum
        of the escaping radiation is 
$$
	\Gamma ~=~ \biggl[ {9 \over 4} ~+~ {\pi^2 m_{\rm e} c^2 \over 
		3 \, (\tau + 2/3)^2 \, kT_{\rm e}} \biggr]^{1/2} ~-~ {1 \over 2}
$$
	(Sunyaev \& Titarchuk 1980; Shapiro et al. 1976).}

\section{Classes of sources of X-ray emission}

We briefly summarize the essential quantities needed to characterize each
component of a SB spectrum:

\noindent
{\bf a) O stars.}                                                   \newline
\noindent
{\it Luminosity:} $L_{2-10} \sim 3 \times 10^{32-33}$ erg s$^{-1}$. \newline
\noindent
{\it Spectrum:} multi-T thermal ($kT \leq 0.5$ keV).                \newline
\noindent
(The contribution of O stars is very minor and will not be
considered further here; see PR02.)

\noindent
{\bf b) Supernova Remnants (SNRs).}                                              \newline
\noindent
{\it Luminosity:} $10^{36} \leq L_{2-10}/({\rm erg \, s}^{-1}) \leq 10^{37}$.    \newline
\noindent
{\it Spectrum:} thermal with $kT \sim 2$ keV, and $Z \sim Z_\odot$.               

\noindent
{\bf c) High-Mass X-Ray Binaries (HMXBs).}                                              \newline
\noindent
{\it Luminosity:} $10^{37} \leq L_{2-10}/({\rm erg \, s}^{-1}) \leq 2 \times 10^{38}$.  \newline
\noindent
{\it Spectrum} in 2-50 keV band:
$$
f(\epsilon) ~ \propto ~ 
\left\{
\begin{array}{ll} \epsilon^{-\gamma}  
& \mbox{ $\,\,\, \epsilon \leq \epsilon_c$} \\
\epsilon^{-\gamma}~ e^{-[(\epsilon-\epsilon_c) / \epsilon_F]}   
& \mbox{ $\,\,\, \epsilon > \epsilon_c$ } 
\end{array} 
\right.
$$
where $\epsilon$ denotes energy (in keV), and with photon index $\gamma \sim1.2$, 
cutoff energy $\epsilon_c \sim 20$ keV, and $e$-folding energy $\epsilon_F \sim 12$ keV. 

\noindent
{\bf d) Very Luminous X-Ray Sources (VLXs).}                                            \newline
\noindent
{\it Luminosity:} $2 \times 10^{38} \leq L_{2-10}/({\rm erg \, s}^{-1}) \leq 10^{39}$.  \newline
\noindent
{\it Spectrum} in 2-10 keV band: power-law with (photon) index $\Gamma \sim 2$.            

\noindent
{\bf e) Ultra-Luminous X-Ray Sources (ULXs).}                                           \newline
\noindent
{\it Luminosity:} $10^{39} \leq L_{2-10}/({\rm erg \, s}^{-1}) \leq 5 \times 10^{39}$.  \newline
\noindent
{\it Spectrum} in 2-10 keV band: power-law with (photon) index $\Gamma \sim 1.3$.          

\noindent
{\bf f) Compton scattering.}              \newline
\noindent
{\it Luminosity:} such that its flux at 10 keV is 0.1 of the total flux (see
discussion in Persic \& Rephaeli 2003).   \newline
\noindent
{\it Spectrum:} power-law with (photon) index $\beta \sim 1.8$.

\section{Luminosity function}

The population of VLX and ULX sources in a galaxy is more abundant the more
active is star formation in that galaxy. Specifically, if the point source
integral luminosity function (XPLF) is described as $N(>L) \propto L^{-\alpha}$
(which corresponds to a differential XPLF of the type ${dN \over dL} \propto
L^{-(1+\alpha)}$), based on a wealth of {\it Chandra} observations it has been found
that XPLFs are less steep in SFGs than in normal spirals and ellipticals.
(For example, the two most nearby SB galaxies, have $\alpha \sim 0.5$ (M82) and 
$\alpha \sim 0.8$ (NGC253), while normal spirals have $\alpha \sim 1.2$, and 
ellipticals have approximately $\alpha \geq 1.4$; see Kilgard et al. 2002.) So 
the integrated point source luminosity of SFGs is dominated by the highest-$L$ 
sources (Colbert et al. 2003).

\section{Synthetic spectrum}

Given an (observed) XPLF, with the "stellar" component clearly sorted by 
luminosity, we can formally compute the integrated "stellar" spectrum 
according to
$$
\phi(\epsilon) ~=~ {\int ~ {dN(L) \over dL}~ L~ f(\epsilon, \,L) ~dL \over 
\int ~ {dN(L) \over dL}~  L~ dL } \,,
$$
where $f(\epsilon, \,L)$ are the above spectral profiles, normalized to the energy
band $\epsilon_1$-$\epsilon_2$ (here, 2-100 keV), where the luminosity $L$ is defined. To 
this we add a diffuse Compton component. To be specific, we choose a power-law
index $\alpha = 0.7$ for the XPLF (i.e. roughly the average between the values
observed for M82 and NGC253), and a Compton component which is 10\% of the 
total flux at 10 keV. It is important to realize that our extrapolation of the 
observed spectra to 100 keV is just for the sake of computation.

\section{Discussion}

The resulting 2-100 keV spectrum is shown in Fig.1. In the synthetic spectrum 
a conspicuous knee, due to the break in the HMXB spectrum, is immediately 
recognizable at 20 keV. If the ULX component, which turns out to lie very close 
to that of HMXBs for energies up to 20 keV, did turn out to have a similar break, 
then the knee in the synthetic spectrum would be sharper. 

\begin{figure}
\vspace{5.0cm}
\includegraphics{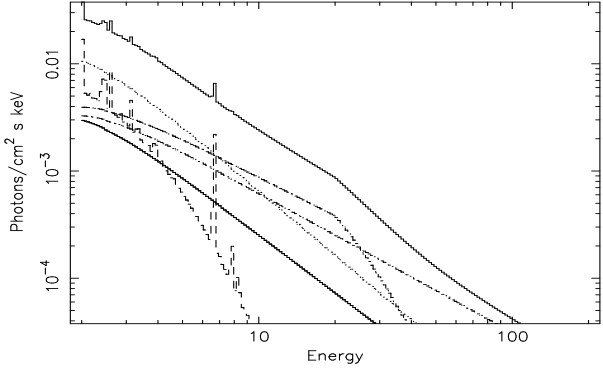}

\caption{The integrated 2-100 keV spectrum of
a starburst. Shown in ascending order at 5 keV are the SNR, Compton, ULX,
HMXB, and VLX components. The "stellar" components are normalized in energy
flux in the 2-100 keV band. The Compton component has been normalized at 0.1
of the total flux at 10 keV. The spectrum is photoelectrically absorbed
through an HI column density $N_{\rm H}= 10^{22}$ cm$^{-2}$.}
\end{figure}

An important related consideration is that the Compton component will be swamped 
by the VLX and ULX components and thus practically undetectable by INTEGRAL, 
unless both "stellar" components exhibit a spectral break similar to that of HMXBs. 
It can be argued that the HMXB spectral break is related to the magnetic field 
strength, $B$, on the surface of the accreting neutron star: 
i) statistically, the cutoff energy, $\epsilon_c$, is correlated with the observed 
cyclotron resonance energy, $\epsilon_0$, by $\epsilon_0 \simeq 2\, \epsilon_c$: as 
the latter depends on $B$ so will the former, which will then provide a measure of 
$B$ (a most helpful feature in those cases where a resonant scattering feature is 
not observed -- White et al. 1995); and 
ii) physically, the spectral break could be due to radiative transfer effects 
in a strong magnetic field (Boldt et al. 1976). If the magnetic fields able 
to induce $\sim$20 keV spectral breaks in VLX-ULX spectra have strengths 
comparable to those observed in HMXBs
	\footnote{A crude estimation suggests that for 
	BH masses in the range $2-8 \,M_\odot$, which 
	can be produced via ordinary evolution (i.e. 
	through type II SN events) by progenitor stars 
	with $M \sim 8-40\, M_\odot$, the conservation of
	magnetic flux implies that for $B \sim 10-100$ G 
	at the surface of the progenitor star, it will be 
	$B \sim 10^{12} - 10^{13}$ G at the Schwarzschild 
	radius, nearly irrespective of the BH mass. (See 
	Maeder \& Meynet 1989 for the relevant quantities
	of the stellar progenitors). These values are 
	consistent with those deduced for the magnetized 
	neutron stars which are members of HMXB systems 
	(see Boldt et al. 1976; White et al. 1995). 
	}, 
then a $\sim$20 keV cutoff is expected also in the spectra of VLX and ULX 
sources. Hence, the knee of the synthetic spectrum will turn out more marked 
and the Compton component will possibly stand out more clearly (at energies
of approximately $\geq$30 keV). INTEGRAL measurements can possibly determine 
whether there is a break in VLX/ULX spectra.

On the other hand the Compton emission, in order to be discernible even in the 
absence of VLX/ULX spectral breaks, has to be stronger than assumed here. In our 
calculation we have assumed the Compton flux to be 0.1 of the total at 10 keV
(which implies that it is $\sim$10$\%$ of the total 2-10 keV flux). This is a 
factor of $\sim$2 higher than the upper limit set by Griffith et al. (2000)
on the Compton flux fraction of M82 observed with {\it Chandra} (see Persic \& 
Rephaeli 2003). It is possible that in SB galaxies high SN rates and dust 
temperatures enhance the Compton component such that it dominates the spectrum 
at energies higher than $\sim$30 keV.


\begin{thebibliography}{}
\bibitem{}
Boldt E.A., Holt S.S., Rothschild R.E., Serlemitsos P.J., 1976, A\&A 50, 161

\bibitem{}
Colbert E.J.M., Heckman T.M., Ptak A.F., Strickland D.K., 2003, ApJ in press (astro-ph/0305476)

\bibitem{}
Fabbiano G., White N.E., 2003, in "Compact Stellar X-Ray Sources", eds.
            W. Lewin \& M. van der Klis (Cambridge University Press) (astro-ph/0307077)

\bibitem{}
Goldshmidt O., Rephaeli Y., 1995, ApJ 444, 113

\bibitem{}
Griffiths R.E., Ptak A., Feigelson E.D., et al., 2000, Science 290, 1325

\bibitem{}
Kilgard R.E., Kaaret P., Krauss M.I., et al., 2002, ApJ 573, 138

\bibitem{}
Liu J.-F., Bregman J.N., Seitzer P., 2002, ApJ 580, L31

\bibitem{}
Maeder A., Meynet G., 1989, A\&A 210, 155

\bibitem{}
Persic M., Rephaeli Y., 2002, A\&A 382, 843 (PR02)

\bibitem{}
Persic M., Rephaeli Y., 2003, A\&A 399, 9

\bibitem{}
Rephaeli Y., 1979, ApJ 227, 364

\bibitem{}
Rephaeli Y., Gruber D., Persic M., 1995, A\&A 300, 91

\bibitem{}
Rephaeli Y., Gruber D., Persic M., McDonald D., 1991, ApJ 380, L59

\bibitem{}
Shapiro S.L., Lightman A.P., Eardley D.M., 1976, ApJ 204, 187

\bibitem{}
Sunyaev R.A., Titarchuk L.G., 1980, A\&A 86, 121

\bibitem{}
White N.E., Nagase F., Parmar A.N., 1995, in 'X-Ray Binaries' (ed. Lewin W.H.G. et al.),
     Cambridge U. Press, 1

\bibitem{}
Zezas A.L., Fabbiano G., Rots A.H., Murray S.S., 2002, ApJ 577, 710
\end{thebibliography}
\end{document}